\documentclass[aps,prb,twocolumn,showpacs]{revtex4-2}
\usepackage{amssymb}
\usepackage{graphicx}
\usepackage{dcolumn}
\usepackage{bm}
\usepackage{amsmath}
\usepackage{soul,color}
\usepackage{textcomp}
\usepackage{times}

\definecolor{cGreen}{RGB}{0,0,0}
\definecolor{cBlue}{RGB}{45,51,180}
\definecolor{cmagenta}{RGB}{205,0,100}

\usepackage[colorlinks,linkcolor=red,citecolor=cBlue]{hyperref}

\begin{document}
	
\title{Imaginary-temperature zeros for quantum phase transitions}
\author{Jinghu Liu$^{1}$}
\author{Shuai Yin$^{2}$}
\email{yinsh6@mail.sysu.edu.cn}
\author{Li Chen$^{1}$}
\email{lchen@sxu.edu.cn}
	
\affiliation{
$^{1}${Institute of Theoretical Physics, State Key Laboratory of Quantum Optics and Quantum Optics Devices, Shanxi University, Taiyuan 030006, China} \\
$^{2}${School of Physics, Sun Yat-Sen University, Guangzhou 510275, China}
}

\begin{abstract}
While the zeros of complex partition functions, such as Lee-Yang zeros and Fisher zeros, have been pivotal in characterizing temperature-driven phase transitions, extending this concept to zero temperature remains an open question. In this work, we propose a solution to this issue by calculating the \textit{imaginary-temperature zeros} (ITZs), which are defined as the roots of the imaginary-temperature partition function.
We illustrate the analytical properties of ITZs in the transverse-field Ising chain, showing that the ITZs' distribution can distinguish between various phases and signify the critical exponents. Universal singular behaviors manifest in such quantities as the edge density of ITZs and the magnetization, with the scaling exponents remarkably differing from those in Lee-Yang theory.
We further illuminate the consistency between ITZs and the zeros of the spectral form factor, which offers a practical path for the experimental detection of ITZs.

\end{abstract}

 \maketitle
\section{Introduction}
Understanding phase transitions has always been a topic of great interest in modern physics. In classical statistical physics, Lee and Yang pioneered the paradigm characterizing temperature-driven phase transitions by the zeros of the complex-valued partition function, known as Lee-Yang zeros \cite{C. N. Yang1952,T. D. Lee1952}. They extended the symmetry-breaking field $h$ to the complex domain and discovered that Lee-Yang zeros of the ferromagnetic Ising model reside in a unit circle of $x=e^{\beta h}$, with $\beta$ the inverse temperature. In the thermodynamic limit, Lee-Yang zeros approach the real axis. Later, Fisher considered the partition function with a complex temperature $y=e^{\beta J}$ \cite{Fisher1965,W. T. Lu2001}, yielding Fisher zeros, with $J$ the coupling strength of spins. For the Ising model, Fisher zeros lie on two circles \cite{M. E. Fisher1978,S.-Y. Kim2005,S. Katsura1967,H. J. Brascam1974}. Over the years, both Lee-Yang and Fisher zeros have been extensively studied theoretically across various models \cite{B.-B. Wei2012,J. Liu2019,R. A. Blythe2003,I. Bena2005,B.-B. Wei2014}. Experimentally, Lee-Yang zeros have been observed by detecting the quantum coherence of a probe spin coupled to an Ising-type spin bath \cite{X. Peng2015}.

In contrast to classical transitions, quantum phase transitions (QPTs) are characterized by nonanalytic changes in ground states at zero temperature and are driven by quantum fluctuations \cite{S. Sachdev1999,S. L. Sondhi1997,M. Vojta2003}. In recent years, QPTs have been elevated to a focal point in both statistical and condensed matter physics \cite{G. Jotzu2014,N. Flaschner2016,I. Bloch2008,M. Lewenstein2007,M. Greiner2002}. As such, understanding QPTs through the lens of generalized partition functions has emerged as a compelling topic. Several studies have already been made in this direction \cite{S. Yin2017, T. Kist2021,P. M. Vecsei2022,P. M. Vecsei2023,G. Y. Chitov2022,P. N. Timonin2021}. Generally, a $d$-dimensional quantum model can be mapped into a $(d+1)$-dimensional classical model \cite{S. Sachdev1999,P. Pfeuty1976,S. Chakravarty1988,A. M. Polyakov1987,S. L. Sondhi1997}. Based on this quantum-classical mapping, both concepts of Lee-Yang zeros and Fisher zeros can be generalized to quantum systems and to characterize QPTs: the former refers to introducing a complex symmetry-breaking term, analogous to the Lee-Yang case \cite{S. Yin2017}; whereas the latter is associated with a complex quantum fluctuation, e.g., the complex transverse field for quantum Ising model \cite{B.-B. Wei2014}. The complex quantum fluctuation serves as the counterpart of Fisher's complex inverse temperature according to the quantum-classical mapping.

Nevertheless, applying Lee-Yang and Fisher's methodologies to quantum cases encounters notable complications. On the one hand, obtaining the relationship between the Lee-Yang or Fisher zeros and QPTs is typically complicated, involving the calculation techniques of the moment-generating functions or tensors \cite{T. Kist2021,P. M. Vecsei2022,P. M. Vecsei2023}. Furthermore, adopting complex magnetic fields breaks the Hermiticity of the system \cite{S. Yin2017, B.-B. Wei2012,X. Peng2015,B.-B. Wei2014}. This imposes additional demands on the system, such as controllable dissipation \cite{C. M. Bender2007}, rendering its experimental realization a challenging task.

Here, we propose a method to characterize QPTs, which is anchored in the \textit{imaginary-temperature partition function}
\begin{equation}
Z(iT,\lambda) = \text{Tr} \left[e^ {- \frac{H(\lambda)}{iT} }\right],
\label{ZT}
\end{equation}
where $H$ is the Hamiltonian, $\lambda$ is a parameter that can trigger a QPT, and $iT$ indicates the imaginary temperature. We focus on the zeros of $Z(iT,\lambda)$. At first glance, our extension seems akin to Fisher's extension on the inverse temperature. However, this is not the case. According to the quantum-classical mapping, Eq.~(\ref{ZT}) represents the imaginary generalization of the extra dimension, thereby featuring a different physical meaning from Fisher's framework. To distinguish it from the latter, we term the zeros of $Z(iT,\lambda)$ as the \textit{imaginary-temperature zeros} (ITZs).

It is also noteworthy to clarify the distinction between the ITZ and the dynamical quantum phase transition (DQPT) \cite{M. Heyl2013,M. Heyl2018}, with the latter being defined as the zeros of the Loschmidt echo \( L(t) = \left\langle \Psi_0 \right| e^{-i H_\text{f} t} \left| \Psi_0 \right\rangle \), where \( \left| \Psi_0 \right\rangle \) denotes the initial state and $H_\text{f}$ is the post-quench Hamiltonian. Thereby, DQPTs are dependent on the initial state. In contrast, our imaginary-temperature partition function [defined in Eq.~(\ref{ZT})] relies solely on the information of the Hamiltonian (i.e., the energy spectrum). This thus allows the ITZs to reflect solely the properties of the Hamiltonian.

We take the transverse-field Ising (TFI) chain as an illustrative example, wherein the solvability of the model allows for an analytical elucidation of ITZs' properties. We will show the following:
(i) In Sec.~\ref{Sec_ITZ}, we illustrate how the distribution of ITZs across different phases can effectively probe the quantum phase transition.
(ii) In Sec.~\ref{Sec_Singularity}, universal singular behaviors are observed in both the ITZs' density and the transverse magnetization.
(iii) In Sec.~\ref{Sec_Critical}, the longitudinal magnetization is shown to be able to signify the Ising critical exponents.
(iv) In Sec.~\ref{Sec_SFF}, we show the partition function $Z$ carries a quantum dynamical interpretation and aligns with the spectral form factor.
Thereafter, in Sec.~\ref{Sec_Application}, we will explore the application of ITZs to other models.
{Finally, a conclusion is given in Sec.~\ref{Sec_Conclusion}.}

\section{Imaginary-Temperature Zeros} \label{Sec_ITZ}
The TFI model under open boundary condition (OBC) is with Hamiltonian
\begin{equation}
H = - \sum_{j=1}^{N-1} \sigma_{j}^z \sigma_{j+1}^z - \lambda \sum_{j=1}^N \sigma_{j}^x,
\label{TFI}
\end{equation}
where $\sigma_{j}^{\{x,z\}}$ represent the spin-$1/2$ Pauli operators, and $N$ is the chain length. The first term accounts for the nearest spin-spin interaction with unit strength, and the second term characterizes the transverse-field strength $\lambda$. The model undergoes a second-order QPT at $\lambda=1$ \cite{S. Sachdev1999,P. Pfeuty1970, A. Dutta2015}. For $\lambda<1$, the ground states are ferromagnetic with twofold degeneracy; for $\lambda>1$, it is paramagnetic without degeneracy.

\begin{figure}[t]
	 \includegraphics[width=0.46\textwidth]{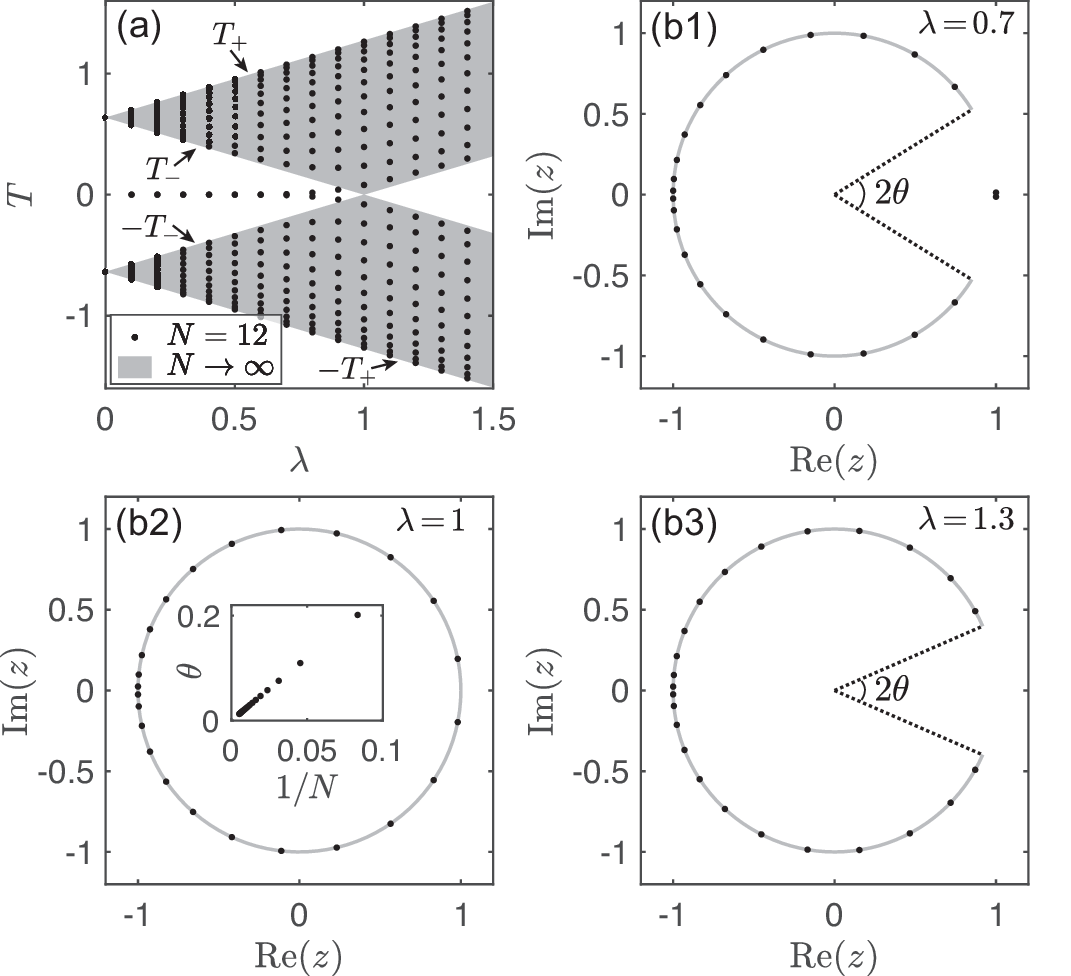}
	\caption{(a) ITZs of the TFI model in the $m=1$ sector. Dots correspond to $N=12$, and the shading indicates the area where ITZs exist in the thermodynamic limit. (b) ITZs of various phases are distributed on the complex $z$ plane, where dots and shading lines correspond to the results for $N=12$ and $N\rightarrow \infty$. The inset of (b2) displays the scaling of $\theta$ on $1/N$. }
	\label{Fig1}
\end{figure}

The quasiparticle spectrum can be obtained by the help of Jordan-Wigner (J-W) and Bogoliubov transformations, with the diagonalized Hamiltonian
$H_\text{f} = \sum_q\epsilon_q(2\eta^\dagger_q\eta_q-1)$,
where $\eta_{q}$ denotes the quasiparticle operator, and $\epsilon_q$ is the quasiparticle spectrum given by
\begin{equation}
\epsilon_{q} = \sqrt{1+\lambda^2+2\lambda \cos q},
\label{HTFdisp}
\end{equation}
which is gapless for $\lambda=1$, and gapped otherwise.
For finite $N$, $q_j$ are determined by the transcendental equation $\sin[(N+1)q]/\sin(Nq)=-\lambda^{-1}$: for $\lambda>1$, $N$ real roots exist; for $\lambda<1$, there are $N-1$ real $q_j$ and a complex $\pi$-mode $q_N \equiv \pi + i q'$ with energy
$\epsilon_{q_N} = \sqrt{1+\lambda^2-2\lambda \cosh q'}$ \cite{P. Pfeuty1970,G. G. Cabrera1987,Y. He2017}.
In the thermodynamic limit, $\epsilon_{q_N} \sim \lambda^N(1-\lambda^2)\rightarrow 0$, indicating the existence of Majorana zero states \cite{K. Chhajed2021,N. Seiberg2023}. More calculation details can be found in Appendix \ref{TFI_OBC}.

We calculate the ITZs based on Eq.~(\ref{ZT}), which takes all the many-body eigenenergies $E_n$ into account. According to $H_\text{f}$, the energies can be expressed as $E_n = \sum_q s_q^n \epsilon_q$, with $s_q^n = \{1,-1\}$ indicating the occupation or vacuum of the $q$ mode. The ground state is with $s_q^0 = -1$ $\forall q$ and $E_0 = -\sum_q \epsilon_{q}$. Then, the partition function can be carried out as
\begin{equation}
Z(iT,\lambda) = \sum_{n} e^{-\frac{\sum_q s_q^n \epsilon_q}{iT}} = 2^N \prod_q \cos\left(\frac{\epsilon_q}{T}\right).
\label{ZTTFI}
\end{equation}
Equation~(\ref{ZTTFI}) reveals that an ITZ would appear when any $q$-mode satisfies $\epsilon_q/T^* = \pm m \pi/2$, or equivalently,
\begin{equation}
T^* = \pm\frac{2\epsilon_q}{ m\pi} = \pm\frac{2\sqrt{1+\lambda^2+2\lambda \cos q}}{ m\pi},
\label{Tstar}
\end{equation}
where $m = \{1,3,5,...\}$ is an odd integer. Note that $m$ divides the zeros into distinct \textit{zero sectors}. Within each sector, there are $2N$ ITZs: $N$ for $T>0$, and $N$ for $T<0$. The distribution of ITZs is symmetric about the $T=0$ axis.

\begin{figure}[t]
\includegraphics[width=0.40\textwidth]{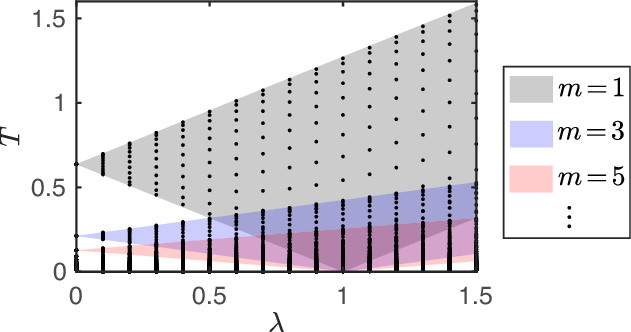}
\caption{An overview of ITZs in the $\protect\lambda$-$T$ plane within $T\ge 0$. Discrete dots
are obtained through numerical calculations with $N=12$, and the shading
indicates the bounded area where ITZs exist in the thermodynamic limit $%
N\rightarrow\infty$. The bounded regions for sectors $m\leq5$ have been
shaded.}
\label{FigS1}
\end{figure}

In Fig.~\ref{Fig1}(a), we illustrate the first zero sector ($m=1$) in the $\lambda$-$T$ plane, where dots represent numerical results for a finite system with $N = 12$. According to Eq.~(\ref{Tstar}), all other zero sectors with $m\neq1$, shown in Fig.~\ref{FigS1}, would exhibit an identical structural pattern to that of the first sector, albeit rescaled by a factor $1/m$ along the $T$ axis.
Since $\cos q \in[-1,1]$, $T^*$ with respect to the real-$q$ modes are bounded within edges. Specifically for $T>0$, the edges are figured out as
\begin{equation}
T_\pm = \frac{2|1\pm \lambda|}{m\pi},
\label{boundary}
\end{equation}
where $\pm$ respectively denote the upper and lower edges, as shown in Fig.~\ref{Fig1}(a). For $T<0$, the edges pick up a minus sign. In the thermodynamic limit where the real-$q$ becomes continuous, the ITZs are continuously distributed within the range $T^*\in[-T_+, -T_-]\cup[T_-, T_+]$, as indicated by shading areas in Fig.~\ref{Fig1}(a).

In the paramagnetic phase $\lambda >1$, all ITZs fall into the bounded regions. At the critical point $\lambda=1$, the lower edges of all sectors converge to the horizontal axis $T_-=0$, serving as a hallmark of the QPT. The convergence is attributable to the gap closing of the bulk spectrum [Eq.~(\ref{HTFdisp})]: a gapless $\epsilon_q$ results in an ITZ at $T^*=0$, as per Eq.~(\ref{Tstar}). In the ferromagnetic phase $\lambda<1$, there exist two isolated lines of ITZs proximate to the horizontal axis, which arises from the complex $\pi$-mode $q_N$. As $N\rightarrow\infty$, these ITZs precisely align with the horizontal axis, owing to $\epsilon_{q_N}\rightarrow0$. Therefore, the unique features of ITZs explicitly diagnose various phases.

In a manner analogous to the aforementioned Lee-Yang circle and Fisher's circle, our ITZs can also be formulated on a unit circle $z \equiv e^{ i \frac{2 \pi T}{w}}$,
where $w \equiv 2T_+$ is the width of the ITZs distribution along the $T$ axis. Figure~\ref{Fig1}(b) displays the ITZs in the complex $z$ plane for various phases. For \(\lambda \neq 1\), the circle is open, with edges $\pm \theta$ corresponding to the lower edges $\pm T_-$ in Fig.~\ref{Fig1}(a). Particularly for the ferromagnetic phase [Fig.~\ref{Fig1}(b1)], isolated ITZs exist within the opening. At the critical point $\lambda = 1$ [Fig.~\ref{Fig1}(b2)], the edges tend to close. The inset presents the scaling of $ \theta$ on $1/N$, which clearly illustrates the closing behavior of edges in the thermodynamic limit.

For a TFI model with the periodic boundary condition (PBC), the Hamiltonian $H$ is translational invariant. In momentum space, the Hamiltonian is decomposed into the independent odd (o) and even (e) channels, with the discrete momenta {$q \in Q$} taking different values:
\begin{equation}
\begin{aligned}
Q_{\mathrm{o}} &=\left\{-\frac{N-2}{N} \pi, \cdots,-\frac{2}{N} \pi, 0, \frac{2}{N} \pi, \cdots, \pi\right\}, \\
Q_{\mathrm{e}} &=\left\{-\frac{N-1}{N} \pi, \cdots,-\frac{1}{N} \pi, \frac{1}{N} \pi, \cdots, \frac{N-1}{N} \pi\right\}.
\end{aligned}
\label{Q}
\end{equation}
The momenta in the two channels are overall offset by $\pi/N$. Similar to the OBC case, the Bogoliubov transformation can recast the Hamiltonian into the diagonal form, i.e., $H_\text{f}=\sum_{q\in Q}\epsilon_{q}\left( 2\eta_{q}^{\dagger}\eta_{q}-1\right)$, where the quasiparticle spectrum $\epsilon_{q}$ is the same to Eq.~(\ref{HTFdisp}).

Then, we can derive the partition function $Z(iT,\lambda)$. The calculation procedural is quite similar to that for the OBC case, but is conducted in the even (odd) channel separately. Eventually, we have
\begin{equation}
\begin{aligned}
Z&=  2^{N-1}\left\{\prod_{q \in Q_{\text {e }}} \cos \left(\frac{\epsilon_q}{T}\right)+\prod_{q \in Q_{\text {o }}} \cos \left(\frac{\epsilon_q}{T}\right) \right. \\
& +\prod_{q \in Q_{\text {e }}}\left[i \sin \left(\frac{\epsilon_q}{T}\right)\right]\left. -\prod_{q \in Q_{\text {o }}}\left[i \sin \left(\frac{\epsilon_q}{T}\right)\right] \right\},
\end{aligned}
\label{Z2}
\end{equation}
where the terms with $q\in Q_{\text {e}}$ and $q\in Q_{\text {o}}$ respectively denote the contributions from the even and odd channels.
It is clearly shown that the partition function is a sum over four {product} terms, which is thus more complex than the OBC’s result [Eq.~(\ref{ZTTFI})]. It is expectable that, in the thermodynamic limit with {$N\rightarrow \infty$, $q$} becomes continuous, the distinctions between odd and even channels would disappear. As a consequence, the last two terms would cancel with each other, and the first two terms would merge with each other, causing Eq.~(\ref{Z2}) to be consistent with the OBC's outcome shown in Eq.~(\ref{ZTTFI}).

\section{Singularities} \label{Sec_Singularity}
One captivating aspect of Lee-Yang theory is the Yang-Lee edge singularity \cite{M. E. Fisher1978}, the universal scaling of Lee-Yang zeros' density $\rho$ near the boundary $h_c$, i.e., $\rho\propto (h - h_c)^\sigma$ with $\sigma$ the exponent. Fisher pointed out that the edge singularity is a critical phenomenon described by a field theory with imaginary coupling \cite{M. E. Fisher1978}. To date, Yang-Lee edge singularity illuminated areas including high-energy physics \cite{W. Li2023,P. Dimopoulos2022}, non-Hermitian systems \cite{C. M. Bender2007}, first-order phase transitions \cite{F. Zhong2005,F. Zhong2012}, etc. This prompts us to explore the singular phenomena of ITZs.

We identify two singular behaviors: the first one is associated with the density of ITZs, characterized by an exponent $\sigma_1 = -1/2$, in contrast to the well-known Yang-Lee edge singularity $\sigma = -1/6$ for the TFI model \cite{J. L. Cardy1985,C. Binek2001}; the second pertains to the divergence of the transverse magnetization with exponent $\sigma_2 = -1$. Let us elaborate on them.

In the thermodynamic limit, the density of ITZs in the first sector can be derived as
\begin{equation}
\rho =
\left\{
\begin{array}{cl}
\frac{2NT}{\pi\sqrt{|(T^2 - T_+^2)(T^2 - T_-^2)|}} & T \in [T_-, T_+] \cup [-T_+, -T_-] \\
0 & \text{elsewhere}
\end{array}
\right. .
\label{TstarDensity}
\end{equation}
In the derivation of Eq.~(\ref{TstarDensity}), we first introduce the
variable $x = q/\pi$ and rewrite $x$ as a
function of $T$, i.e.,
\begin{equation}
x=\frac{1}{\pi}\arccos\frac{\left( \frac{m\pi T}{2}\right) ^{2}-\left(
\lambda^{2}+1\right) }{2\lambda}.
\end{equation}
Then, the density can be obtained by $\rho(T) =N dx/dT$.
Equation~(\ref{TstarDensity}) indicates that the density diverges $\rho\propto |T-T_\pm|^{\sigma_1}$ at sector edges, with exponent $\sigma_1 = -1/2$. Notably, $\sigma_1$ is independent of $\lambda$, except at the critical point $\lambda =1$.
At the critical point, $\rho(T)$ only diverges at the upper boundary $T_+$, while maintaining a finite density $\rho(0) = N/2$ at the lower boundary $T_-=0$. The latter is confirmed by

\begin{figure}[t]
 \includegraphics[width=0.48\textwidth]{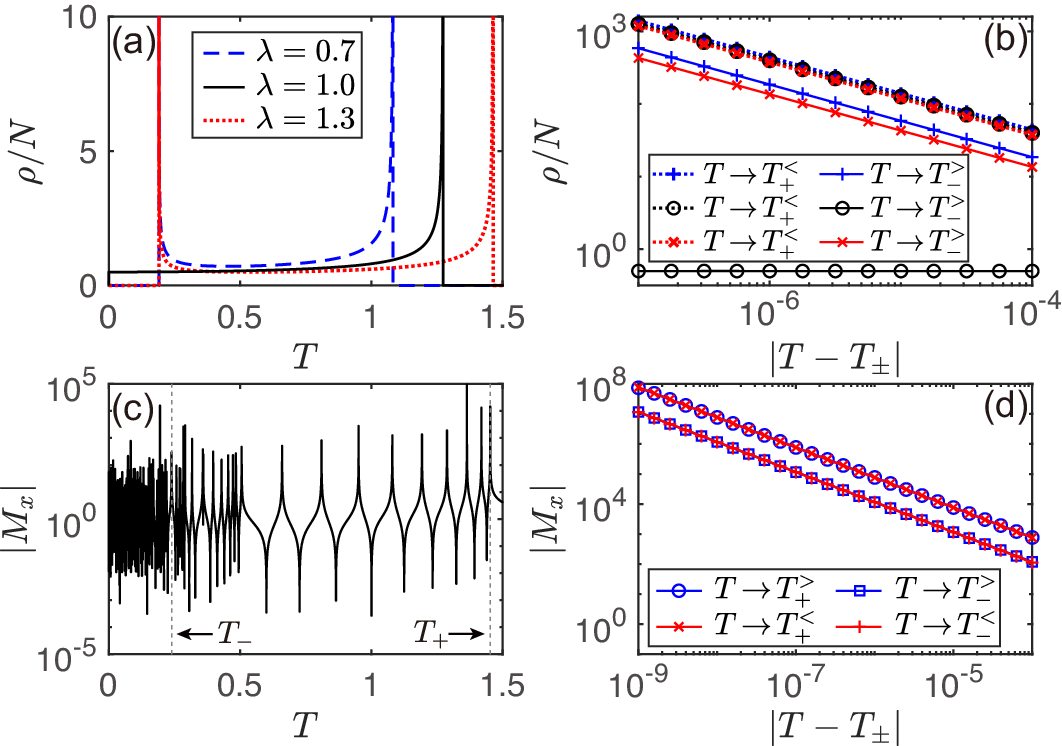}
	\caption{(a) ITZs' density $\rho(T)$ of the $m=1$ sector within $T\ge 0$. (b) Scaling behavior of $\rho(T)$ near the sector edges $T = T_\pm$, where $T^<$ and $T^>$ denote approaches to the edges from the left and right, respectively. (c) Transverse magnetization $|M_x|$ as a function of $T$, with dotted lines indicating the edge locations of the $m=1$ sector. (d) Scaling of the singularities in $|M_x|$ near the edges.}
	\label{Fig2}
\end{figure}

\begin{equation}
\lim_{T\rightarrow0}\rho(T)  =\lim_{T\rightarrow0}\frac
	{2N}{\pi\left\vert T^{2}-T_{+}^{2}\right\vert^{1/2}}
	=\frac{N}{2}.
\label{TstarDensityLimit}
\end{equation}
In Fig.~\ref{Fig2}(a), we plot $\rho$ for the cases $\lambda=0.7$, $1$, and $1.3$. Furthermore, Fig.~\ref{Fig2}(b) shows the scaling behaviors of $\rho$ near the edges $T = T_\pm$ for the corresponding $\lambda$ of Fig.~\ref{Fig2}(a).

Next, we turn to the magnetization defined as [according to Eq.~(\ref{ZT})]
\begin{equation}
M_{x,z}(T) = \frac{\text{Tr} \left[ e^{-\frac{H}{iT}} S_{x,z} \right]}{Z(iT,\lambda)} = \frac{\sum_ne^{-\frac{E_n}{iT}}\langle n|S_{x,z} |n\rangle}{Z(iT,\lambda)},
\label{MT}
\end{equation}
where $S_{x,z}= N^{-1}\sum_{j} \sigma_j^{x,z}$. In contrast to the Lee-Yang theory which is symmetry breaking \cite{M. E. Fisher1978}, our longitudinal magnetization $M_z$ is always vanishing since $Z(iT,\lambda)$ respects the spin-flip symmetry. For the transverse magnetization $M_x$, the numerator of Eq.~(\ref{MT}) is smooth and differentiable for $T>0$; likewise, the denominator $Z(iT,\lambda)$ is also a smooth function, albeit with zeros. This suggests that the singular points in $M_x$ should coincide with the ITZs, as shown in Fig.~\ref{Fig2}(c). As $T$ approaches $T^{*}$, the absolute value of $%
M_{x}(T)$ can be expanded as a series in terms of $|T - T^{*}|$:
\begin{equation}
\begin{aligned}
\lim_{T\rightarrow T^{\ast}}\left\vert M_{x}\right\vert &
\propto \lim_{T\rightarrow T^{\ast}}\frac{1}{|Z\left( iT,\lambda\right)| }\\ & =|\frac{2T^{\ast}}{\pi}\left( T-T^{\ast}\right)
^{-1}+\frac{2}{\pi}+O\left[ (T-T^{\ast}) \right] |
\end{aligned}.
\end{equation}
It indicates that $M_x$ would diverge in the way of $|M_x|\propto |T-T^*|^{\sigma_2}$ with $\sigma_2 = -1$ . The scaling component $\sigma_2$ is numerically verified in Fig.~\ref{Fig2}(d).

\section{Relation to Spectra Form Factor} \label{Sec_SFF}
The partition function $Z(iT,\lambda)$ carries an additional layer of physical meaning in quantum dynamics. To elucidate this, we consider the modula
\begin{equation}
\begin{aligned}
|Z(iT,\lambda)|^2 &= | \sum_{n} \langle n | e^{-\frac{H}{iT}}| n \rangle |^2 \\
&= D^2 | \langle \Psi_0 | e^{-i\frac{H}{T}}| \Psi_0 \rangle |^2,
\end{aligned}
\label{ZT2}
\end{equation}
where $| \Psi_0 \rangle = D^{-1/2}\sum_n |n \rangle$, and $D=2^N$ denotes the dimension of Hilbert space.
Clearly, by setting $t \sim 1/T$, $|Z(t)|^2$ represents the return probability when the initial state $| \Psi_0 \rangle$ is an equally weighted superposition of all eigenstates $|n\rangle$. The return probability essentially links to the spectral form factor (SFF) \cite{G. Bunin2022,C. B. Dag2023,J. Suntajs2020,A. Prakash2021,X. Chen2018,M. Winer2022,P. Kos2018,M. Schiulaz2019} in the manner of
\begin{equation}
K(t) = \frac{1}{D} \left\langle \bigg\vert\sum_{E} g(E)e^{-i 2 \pi E t} \bigg\vert^2\right\rangle = \frac{|Z(2\pi t)|^2}{D},
\label{SFF}
\end{equation}
where $g(E)$ is the density of state, and $\langle \cdots \rangle$ denotes the average on a spectral ensemble.
For the current system, the entire spectrum can be precisely solved, rendering the ensemble average irrelevant. Therefore, up to the factor $D$, there is a one-to-one correspondence between the ITZs and the zeros of the SFF, satisfying $t^* = 1/2\pi T^*$.

Commonly, the envelope structure of SFF can be used to distinguish various quantum states. For example, in chaotic systems, the SFF displays a three-stage "slope-ramp-plateau" structure \cite{G. Bunin2022,C. B. Dag2023,J. Suntajs2020,A. Prakash2021,X. Chen2018,M. Winer2022,P. Kos2018,M. Schiulaz2019}, with the two characteristic time scales as the Thouless time and Heisenberg time; the ratio of the two time scales can further be used to assess the ergodicity breaking.
Here, we emphasize that the zeros of $K(t)$ offer important information in signifying quantum criticality.

\begin{figure}[t]
	 \includegraphics[width=0.46\textwidth]{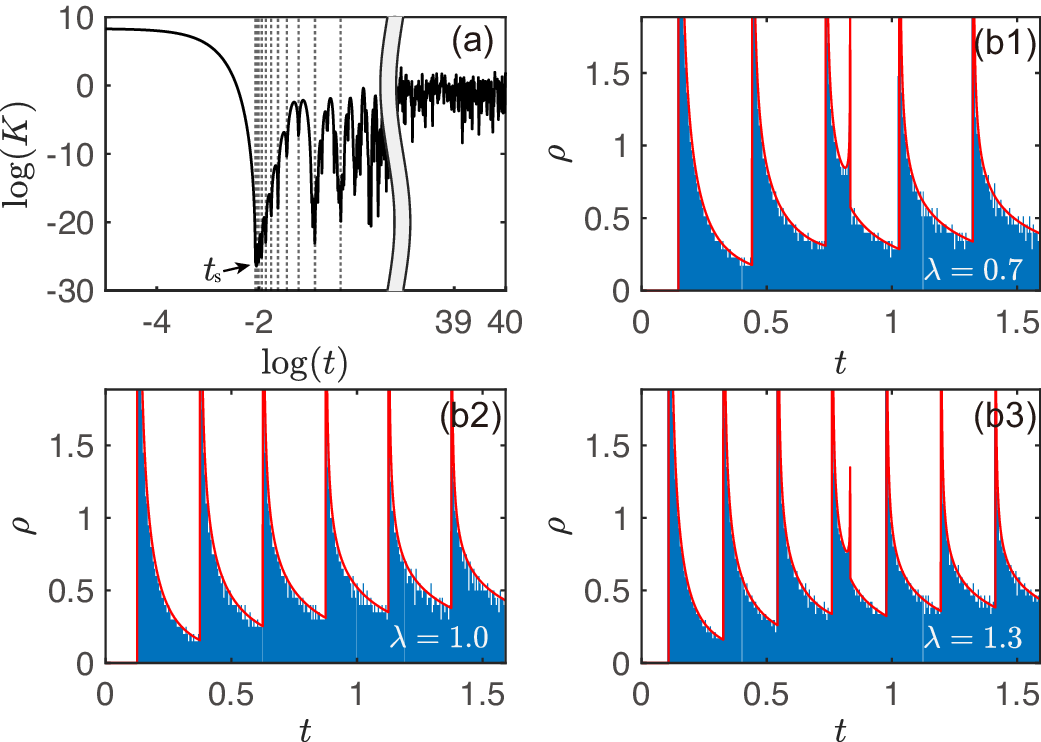}
	\caption{(a) Log-log plot of the SFF for $N=12$ and $\lambda=1$. Vertical dotted lines verify the correspondence between the zeros of $K(t)$ and the ITZs in the $m=1$ sector. (b) Density distribution of the zeros of $K(t)$. Solid lines denote analytical results, while bars are derived from numerical statistics for $N=10^3$. Panels (b1)–(b3) correspond to the cases of $\lambda=0.7$, $1$, and $1.3$, respectively. }
	\label{Fig3}
\end{figure}

In Fig.~\ref{Fig3}(a), we present the numerical result of $K(t)$ for $N=12$ and $\lambda=1$. The plot also reveals a three-stage behavior. Prior to the timescale $t_\text{s} =1/{4(1+\lambda)}$, $K(t)$ declines smoothly devoid of zeros. This stage corresponds to $T$ decreasing from infinity down to the upper boundary $T_+$ of the $m=1$ sector. After $t_\text{s}$, $K(t)$ manifests zeros.
We have verified that the indicated zeros of $K(t)$ [by the vertical dotted lines in Fig.~\ref{Fig3}(a)] strictly correspond to the ITZs of the $m=1$ sector, satisfying $t^* = 1/2\pi T^*$. Other unindicated zeros of $K(t)$ correspond to the ITZs in other sectors. For sufficiently large $t$, $K(t)$ stabilizes at a plateau $\sim 1$, indicating the time scale beyond the Heisenberg time.

Examining the zero-density $\rho(t)$ of the SFF helps to identify the critical point. As mentioned before, at the critical point $\lambda=1$, the density of ITZs only diverges at the upper boundary $T_+$ of all zero sectors. Consequently, $\rho(t)$ manifests periodic peaks, as shown in Fig.~\ref{Fig3}(b2), in a period of $2t_s$. In contrast, for the noncritical cases, additional peaks emerge, as illustrated in Figs.~\ref{Fig3}(b1) and (b3). These additional peaks originate from the singularities at the lower boundaries $T_-$. In Fig.~\ref{Fig3}(b), solid lines delineate the analytical results in the thermodynamic limit based on Eq.~(\ref{TstarDensity}), whereas the bars represent the numerical statistical results for $N=10^3$. Two results are in excellent agreement.

\section{Signature of Critical Exponents} \label{Sec_Critical}

We next turn to the question: can ITZs signify the Ising critical exponents? The short answer is yes.
At first, the shading area in Fig.~\ref{Fig1}(a)  can be treated as a "critical fan". In this sense, in proximity to $\lambda = 1$, the lower edge $T_-$ defines a characteristic energy connected to the dynamic exponent $s$. According to critical theory \cite{S. Sachdev1999,A. Dutta2015}, the characteristic energy should satisfy $T_- \propto (\lambda - 1)^{\nu s}$, where $\nu$ is the correlation length exponent. For the TFI model, $\nu = s=1$, yielding $T_- \propto (\lambda - 1)$, which confirms Eq.~(\ref{boundary}).

\begin{figure}[t]
	 \includegraphics[width=0.48\textwidth]{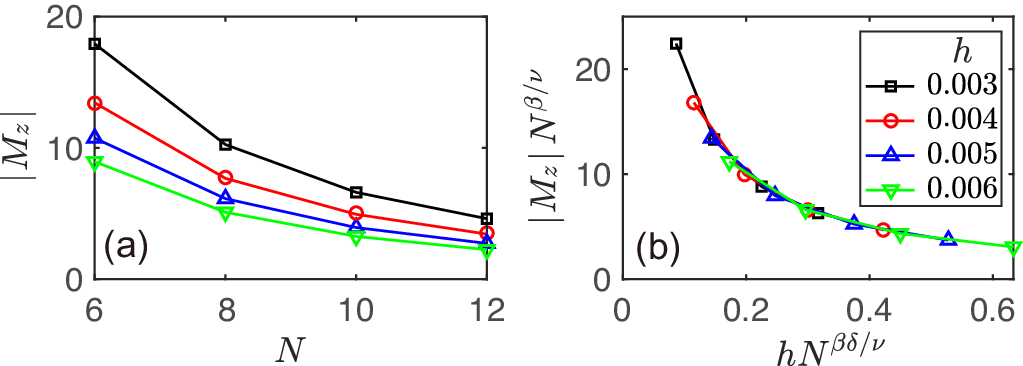}
	\caption{(a) Longitudinal magnetization $|M_z|$ as a function of $N$, with $(\lambda-1)N=0.1$ being fixed. (b) Rescaled plots of $|M_z|$ and $N$. In both panels, various markers denote cases corresponding to different probe field strengths $h$.}
	\label{Fig4}
\end{figure}

Furthermore, probing the system by a weak longitudinal magnetic field $-h \sum_{j} \sigma_j^{z}$ leads to a nonvanishing $M_z$ [defined in Eq.~(\ref{MT})], which is shown in Fig.~\ref{Fig4}(a). We find that, near $\lambda = 1$, the behavior of $M_z$ at the lower edge $T_-$ is determined by the Ising exponents, i.e.,
\begin{equation}
|M_z(T_-)| = N^{-\beta/\nu} f[(\lambda-1)N^{1/\nu},hN^{\beta\delta/\nu},T_-(\lambda-1)^{-\nu s}],
\label{Mz}
\end{equation}
where $f$ is the scaling function, and $\beta=1/8$, $\delta=15$ are Ising exponents of the magnetization and the response exponent to external field, respectively. Therefore, for a fixed $(\lambda-1)N^{1/\nu}$, by rescaling $M_z$ and $N$ into $M_z N^{\beta/\nu}$ and $hN^{\beta\delta/\nu}$, all the curves should match with each other, which is confirmed in Fig.~\ref{Fig4}(b).

\section{Application to other Models} \label{Sec_Application}
So far, we have provided an overview of ITZs' properties based on the TFI model. Now, we briefly discuss the application to some other models.

\subsection{XX model}

\begin{figure}[t]
	 \includegraphics[width=0.46\textwidth]{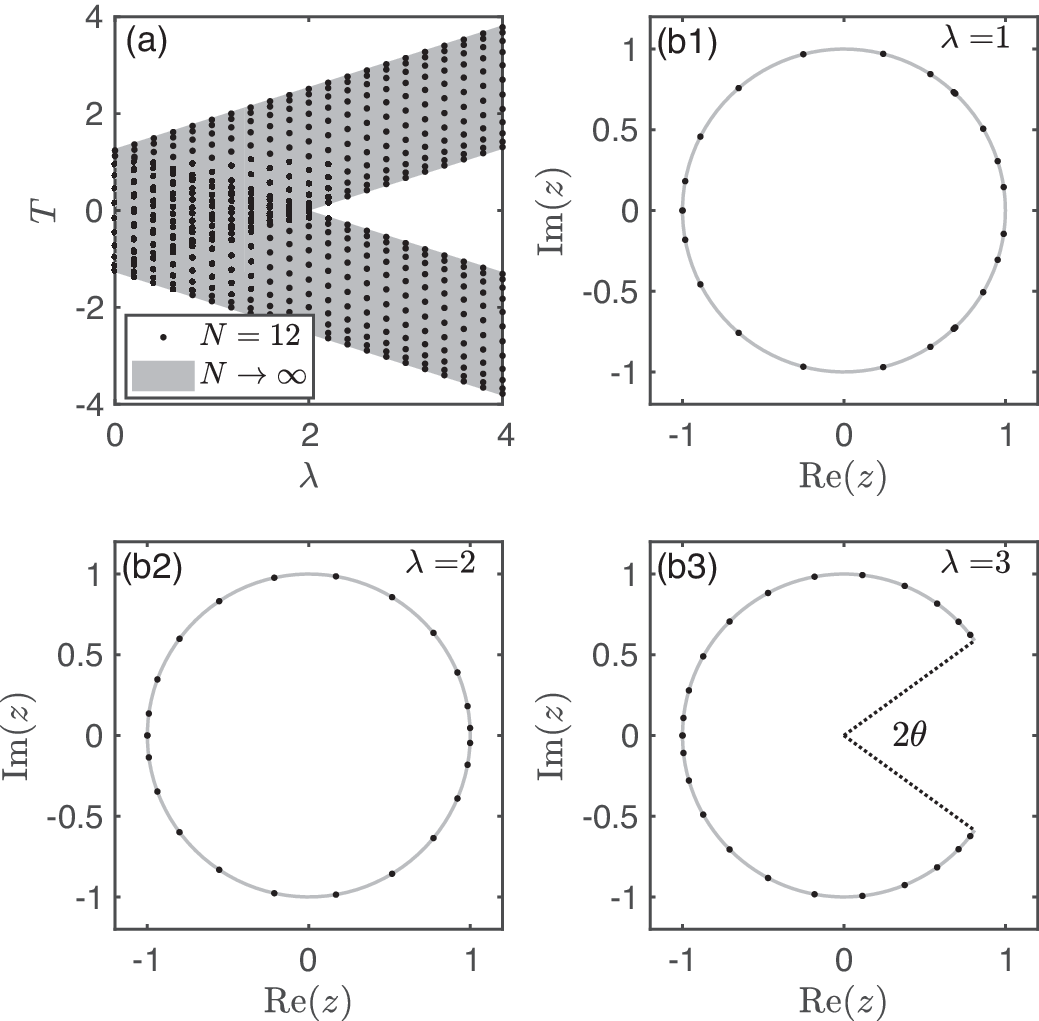}
	\caption{(a) ITZs of the XX model in the first sector. (b) ITZs of various phases distributed on the complex $z$ plane, with (b1)–(b3) corresponding to the cases of $\lambda = 1$, $\lambda = 2$ (multicritical point), and $\lambda = 3$, respectively.
	\label{Fig5}}
\end{figure}

The XX model is solvable and characterized by the Hamiltonian
\begin{equation}
H=-\sum_{i=1}^{N-1}\left( \sigma _{i}^{x}\sigma _{i+1}^{x}+\sigma
_{i}^{y}\sigma _{i+1}^{y}\right)-\lambda \sum_{i=1}^{N}\sigma _{i}^{z},
\label{XX}
\end{equation}
which carries the spin-rotational symmetry along $z$ direction and exhibits a transition from the gapless liquid phase (in $\lambda<2$) to the paramagnetic phase (in $\lambda>2$), with $\lambda = 2$ being called the multicritical point.
Under the J-W transformation, the fermionic Hamiltonian can be diagonalized in the quasiparticle space as
$
H=\sum_{q}\epsilon _{q}\left( 2{\eta }_{q}^{\dagger }{\eta }_{q}+1\right),
$
with the quasiparticle spectrum $\epsilon _{q}$ is figured out to be
\begin{equation}
\epsilon _{q}=\left\vert \lambda +2\cos q\right\vert,
\end{equation}
with $q=\left\{ \frac{\pi }{N+1},\frac{2\pi }{N+1},\frac{3\pi }{N+1}\ldots ,\frac{%
N\pi }{N+1} \right\}$.
Then, following a similar procedure, one can obtain the ITZs as
\begin{equation}
T^{*}=\frac{%
2\left\vert \lambda +2\cos q\right\vert }{m\pi }.%
\end{equation}

The ITZs in the $m=1$ sector for both cases of a finite $N$ and $N\rightarrow \infty$ are shown in Fig.~\ref{Fig5}(a). In the liquid phase, the lower sectoral edge $T_-$ is always zero, in contrast to the ferromagnetic phase of the TFI model. Therefore, when viewed in the complex $z$ plane, as shown in Fig.~\ref{Fig5}(b), the circular zeros exhibit an opening in the paramagnetic phase, while the contour is always closed in the liquid phase and at the multicritical point. On the right side of the multicritical point, the lower edge $T_- \propto (\lambda - 2)^{\nu s}$ remains linear, with $\nu = 1/2$ and $s = 2$ being the critical exponents of the XX model \cite{A. Dutta2015}.

Additional calculations show that, in the thermodynamic limit, the density of ITZs (for $T>0$) behaves as
\begin{equation}
\begin{aligned}
\rho (T)=&\frac{1}{\pi \sqrt{\left( T+\frac{4-2\lambda }{m\pi }%
\right) \left( T_{+}-T\right) }} \\
&+\frac{\Theta (2-\lambda )}{\pi \sqrt{\left(
\frac{4-2\lambda }{m\pi }-T\right) \left( T_{+}+T\right) }},
\end{aligned}
\end{equation}
where $\Theta (x)=1$ for $x>0$ and $\Theta (x)=0$ for $x<0$.
It indicates that $\rho (T)$ also diverges at the gapped sector edges in the exponent $\sigma_1 = -1/2$. Additionally, the transverse magnetization also diverges near the zeros with exponent $\sigma_2 = -1$, i.e.,
\begin{equation}
\begin{aligned}
\lim_{T\rightarrow T^{\ast }}\left\vert M_{x}\right\vert &\propto
\lim_{T\rightarrow T^{\ast }}\frac{1}{|Z\left( iT,\lambda \right) |} \\
&\approx \left |\frac{2T^{\ast }}{\pi }\left( T-T^{\ast }\right)^{-1} \right|.
\end{aligned}
\end{equation}
These divergent behaviors are consistent with those observed in the TFI model.

\subsection{Potts models}

\begin{figure}[t]
\includegraphics[width=0.48\textwidth]{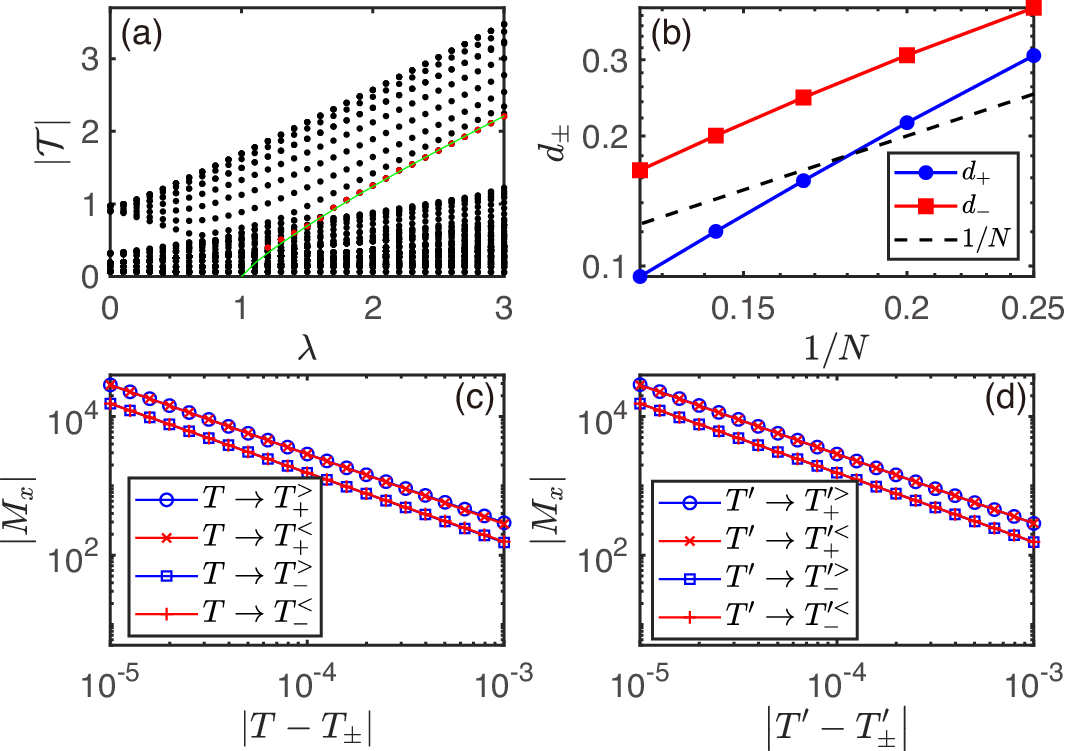}
\caption{Results for the three-state Potts model.
(a) Zeros in the $\lambda$-$|\mathcal{T}|$ plane for $N=8$.
(b) Scaling behaviors of the distance $d$ between two zeros near the sector edges $\mathcal{T}_\pm$.
(c) and (d) display the scaling of $|M_x|$ near the sector edges, with $T_\pm$ and $T'_\pm$ denoting the imaginary and real parts of $\mathcal{T}_\pm$, respectively.
\label{FigS6}
}
\end{figure}

We additionally consider the three-state and four-state quantum Potts models, which are nonintegrable for $\lambda \neq 1$ \cite{ A. Rapp2013, N. Chepiga2022}. The lack of integrability necessitates numerical diagonalization of the Hamiltonian, which is thus limited to systems with a few particles. Generally, the $q$-state Potts model is characterized by the Hamiltonian
\begin{equation}
H=-\sum_j^{N-1} \sum_{d=1}^{q-1} M_j^d M_{j+1}^{q-d}-\lambda \sum_j^N S_j^x,
\label{HPotts}
\end{equation}
where $M$ and $S^x$ are local Potts matrices, being specifically
\begin{equation}
\begin{aligned}
M=\left(\begin{array}{lll}
0 & 1 & 0 \\
0 & 0 & 1 \\
1 & 0 & 0
\end{array}\right), \ \ \
S^x=\left(\begin{array}{ccc}
2 & 0 & 0 \\
0 & -1 & 0 \\
0 & 0 & -1
\end{array}\right),
\end{aligned}
\end{equation}
for $q=3$,
and
\begin{equation}
\begin{aligned}
M=\left(\begin{array}{llll}
0 & 1 & 0 & 0 \\
0 & 0 & 1 & 0 \\
0 & 0 & 0 & 1 \\
1 & 0  & 0 & 0
\end{array}\right), \ \ \
S^x=\left(\begin{array}{cccc}
3 & 0 & 0 &0 \\
0 & -1 & 0 &0 \\
0 & 0 & -1 &0 \\
0 & 0 & 0 &-1 \\
\end{array}\right),
\end{aligned}
\end{equation}
for $q=4$.
The three-state and four-state Potts models exhibit continuous quantum phase transitions from the ordered phase $(\lambda<1)$ to the disordered phase $(\lambda>1)$, with the critical point $\lambda=1$ falling into the three-state Potts and the Ashkin-Teller universality classes, respectively.

We numerically identify the zeros by directly diagonalizing the Hamiltonian [Eq.~(\ref{HPotts})], and find that the zeros now exist in a space spanned by $\lambda$ and $\mathcal{T}$, where $\mathcal{T} = T' + iT$ represents the complex temperature with $T'$ being its real part. In such a space, the zeros' distribution maintains symmetric about the $T = 0$ axis. This phenomenon implies that the zeros can be well defined by the \textit{complex-temperature partition function}
\begin{equation}
Z = \text{Tr}[e^{-\frac{H(\lambda)}{\mathcal{T}}}],
\label{Ztau}
\end{equation}
a further generalization of Eq.~(\ref{ZT}). By doing so, we thus are allowed to discuss the properties of the zeros within the $\lambda$-$|\mathcal{T}|$ plane.

\begin{figure}[t]
\includegraphics[width=0.47\textwidth]{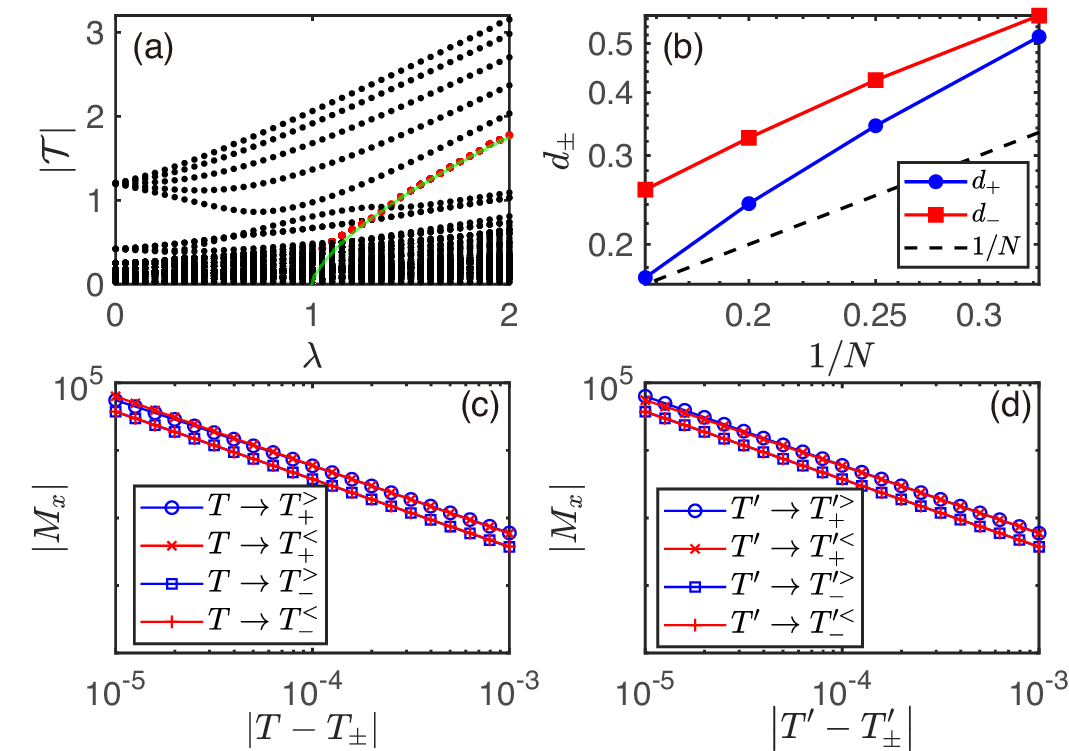}
\caption{Results for the four-state Potts model.
(a) Zeros in the $\lambda$-$|\mathcal{T}|$ plane for $N=6$.
(b) Scaling behaviors of $d$.
(c) and (d) illustrate the scaling of $|M_x|$.
\label{FigS7}
}
\end{figure}

{In Fig.~\ref{FigS6}(a), we show} the identified zeros of the three-state Potts model in $\lambda$-$|\mathcal{T}|$ plane, where the first zero sector is visible. The zeros share very similar properties to those of the TFI model.
The isolated line of zeros, as indicated by the red dots, still exists. These zeros are gapped from $|\mathcal{T}|=0$ in the disordered phase, and tend to approach $|\mathcal{T}|=0$ as $\lambda$ decreases. However, near the horizontal axis, the zeros become too dense such that it becomes challenging to distinguish the isolated line of zeros from other zeros clearly. Hence, Fig.~\ref{FigS6}(a) only markes the zeros that can be confidently identified.

Furthermore, in the disordered phase $\lambda > 1$, the zero line fits well with the function $\propto (\lambda - 1)^{\nu s}$ [as indicated by the green curve in Fig.~\ref{FigS6}(a)], where $\nu = 5/6$ and $s = 1$ are the critical exponents of the three-state Potts model \cite{A. Rapp2013, N. Chepiga2022, F. Y. Wu1982}.
As $N$ increases, the zeros exhibit a gathering behavior towards the sector edges. This can be numerically confirmed by the scaling of $d_\pm$ as a function of $1/N$, as shown in Fig.~\ref{FigS6}(b), where $d_+$ and  $d_-$ represent the distance between two adjacent zeros near the upper and lower edges, respectively. It can be observed that both $d_\pm$ decay faster than $1/N$ [denoted by the dashed line], where the latter represents the distance of a uniform distribution. Hence, near the sector edges, the zero density tends to increase faster than the uniform distribution.
The transverse magnetization $M_x$ near the sector edge also diverges as $(T-T_\pm)^{-1}$ and $(T'-T'_\pm)^{-1}$ [shown in Figs.~\ref{FigS6}(c) and (d)], confirming the exponent $\sigma_2 = -1$.

The four-state Potts model {exhibits} similar features to the three-state Potts, as shown in Fig.~\ref{FigS7}. Now, the zero line can fit with the function $\propto (\lambda - 1)^{\nu s}$ [see Fig.~\ref{FigS7}(a)] in the disordered phase near the critical point, where $\nu = 2/3$ and $s = 1$ are the Ashkin-Teller critical exponents.

\section{Conclusion} \label{Sec_Conclusion}

To conclude, we proposed to signify the QPTs by calculating the zeros of the imaginary-temperature partition function.
Our analysis demonstrated that the distribution of zeros can diagnose different phases; the zeros' edge is capable of reflecting the critical properties.
Both zeros' density and the magnetization exhibit universal singular behaviors.
The link between the zeros of the partition function and the zeros of the spectral form factor has also been unveiled.
Our approach could be potentially extended to the complex-temperature domain for broader applications.
Recently, we were informed of a work on observing the Yang-Lee edge singularity in an encoded photonic system \cite{Gao2024}. The experimental platform fits well with our theoretical proposal and holds promise for direct observations of the zeros.

\begin{acknowledgments}
J. L. and L. C. would like to thank Markus Heyl for the insightful discussion. S. Y. is supported by the NSF of China (Grants No. 12075324 and No. 12222515). L. C. acknowledges support from the NSF of China (Grants No. 12174236) and from the fund for the Shanxi 1331 Project.
\end{acknowledgments}

\appendix

\section{Calculation Details for the TFI Model}
\label{TFI_OBC}

Under the J-W transformation, the TFI model under OBC is recast into a quadratic form
\begin{equation}
H_{f}=N\lambda+\sum_{ij}^{N}\left[ {c}_{i}^{\dagger}A_{ij}{c}_{j}+\frac{1}{2}%
\left( {c}_{i}^{\dagger}B_{ij}{c}_{j}^{\dagger}-{c}_{i}B_{ij}{c}_{j}\right) %
\right] ,  \tag{A1}  \label{H_quadratic}
\end{equation}
where the ${c}_{i}$ and ${c}_{i}^{\dagger}$ are fermionic field operators, $%
A_{ij}=-2\lambda\delta_{i,j}-\left( \delta_{i,j+1}+\delta_{i+1,j}\right)$,
and $B_{ij}=-\left( \delta_{i,j+1}-\delta_{i+1,j}\right) $. Due to
the lack of translational symmetry, $H_{f}$ cannot be diagonalized in
momentum space but can be diagonalized in real quasiparticle space \cite{P.
Pfeuty1970, G. G. Cabrera1987,Y. He2017}. The transformation between
quasiparticle operators and the fermionic operators are given by
\begin{equation}
\begin{aligned} {\eta}_{q}=\sum_{i}\left(
g_{qi}{c}_{i}+\tilde{g}_{qi}{c}_{i}^{\dagger }\right), \\
{\eta}_{q}^{\dagger }=\sum_{i}\left( g_{qi}{c} _{i}^{\dagger
}+\tilde{g}_{qi}{c}_{i}\right) , \end{aligned}  \tag{A2}  \label{g_h}
\end{equation}
where $g_{qi}$ and $\tilde{g}_{qi}$ are real variational parameters. The
quasiparticle operators should satisfy the fermionic anticommutation
relations
\begin{equation}
\{{\eta}_{q},{\eta}_{q^{\prime}}^{\dagger}\}=\delta_{q,q^{\prime}}\text{, }\
\ \{{\eta}_{q},{\eta}_{q^{\prime}}\}=\{{\eta}_{q}^{\dagger},{\eta }%
_{q^{\prime}}^{\dagger}\}=0\text{,}  \tag{A3}
\label{quasi_particle_ancommute}
\end{equation}
which requires the parameters to satisfy
\begin{equation}
\begin{aligned} \sum_{i}\left( g_{qi}g_{q^{\prime
}i}+\tilde{g}_{qi}\tilde{g}_{q^{\prime }i}\right) &= \delta _{q,q^{\prime
}}, \\ \sum_{i}\left( g_{qi}\tilde{g}_{q^{\prime
}i}+\tilde{g}_{qi}g_{q^{\prime }i}\right) &= 0. \end{aligned}  \tag{A4}
\end{equation}
Based on this, the Hamiltonian Eq.~(\ref{H_quadratic}) can be diagonalized
as
\begin{equation}
H=\sum_{q}\epsilon_{q}\left( 2{\eta}_{q}^{\dagger}{\eta}_{q}+1\right) \text{.%
}  \tag{A5}  \label{H_diagonalize}
\end{equation}

\begin{figure}[t]
\includegraphics[width=0.48\textwidth]{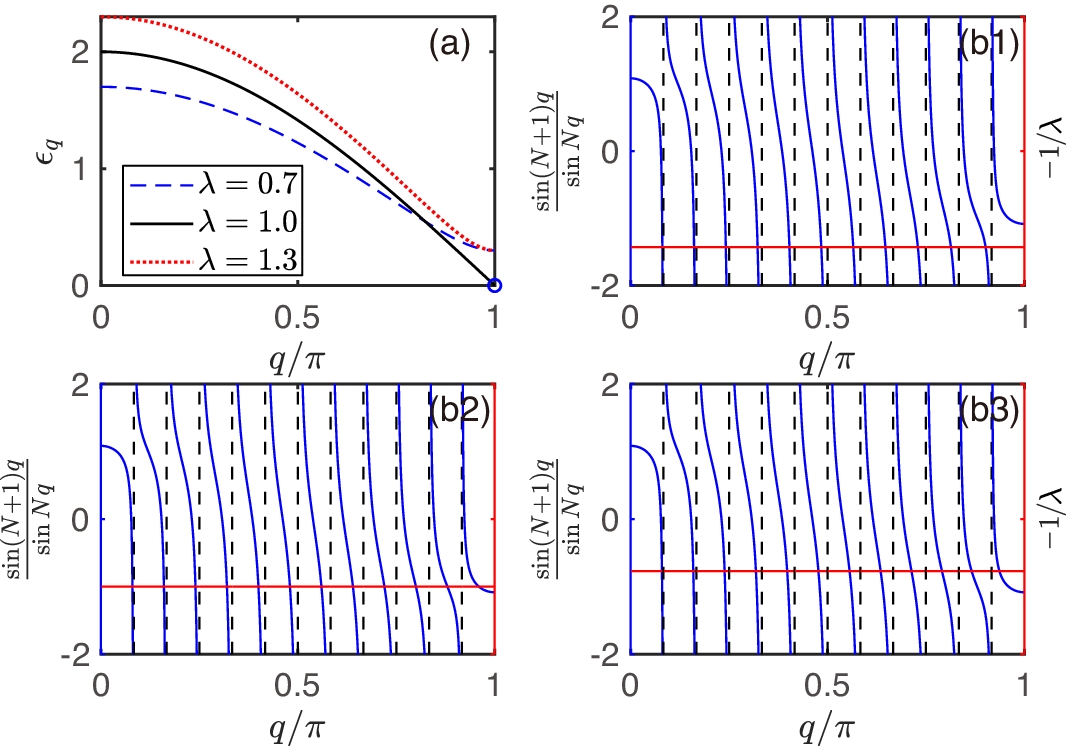}
\caption{(a) Quasiparticle energy $\protect\epsilon_{q}$ for the TFI chain,
where lines and the circle correspond to the real-$q$ modes and the complex-$%
\protect\pi$ mode, respectively. (b) Solutions of the Eq.~(\protect\ref{TEQ}) for $N=12$, with blue and red lines, respectively, indicating the L.H.S. and
R.H.S. of Eq.~(\protect\ref{TEQ}); black dashed lines are the asymptotic
lines with $q_{i}/\protect\pi=i/N$ with $i=1,2,3,...,N-1$. Panels (b1)–(b3)
show cases $\protect\lambda=0.7$, $\protect\lambda=1$, and $\protect\lambda%
=1.3$, respectively. }
\label{Fig_SM2}
\end{figure}

To determine the quasiparticle spectrum $\epsilon_{q}$, we introduce
\begin{equation}
\begin{aligned} \phi _{qi}=g_{qi}+\tilde{g}_{qi}, \\ \psi
_{qi}=g_{qi}-\tilde{g}_{qi}, \end{aligned}  \tag{A6}
\end{equation}
which satisfy $\sum_{q}\phi_{qi}\phi_{qj}=\delta_{ij}$ and $\sum_{q}\psi
_{qi}\psi_{qj}=\delta_{ij}$. Using the relation $[ {\eta}_{q}, {H}%
]=2\epsilon_{q} {\eta}_{q}$, along with Eqs.~(\ref{H_quadratic}) and (\ref%
{g_h}), we arrive at the coupled matrix equations $\left( \mathbf{A+B}%
\right) \mathbf{\phi}_{q}=2\epsilon_{q}\mathbf{\psi}_{q}$ and $\left(
\mathbf{A-B}\right) \mathbf{\psi}_{q}=2\epsilon_{q}\mathbf{\phi}_{q}$. These
equations can be decoupled to yield the eigenvalue equations
\begin{equation}
\begin{aligned} \left( \mathbf{A-B}\right) \left( \mathbf{A+B}\right)
\mathbf{\phi }_{q}=4\epsilon _{q}^{2}\mathbf{\phi }_{q},\\ \left(
\mathbf{A+B}\right) \left( \mathbf{A-B}\right) \mathbf{\psi }_{q}=4\epsilon
_{q}^{2}\mathbf{\psi }_{q}. \end{aligned}  \tag{A7}
\label{coupled_equations3}
\end{equation}

The quasi-particle spectrum $\epsilon_{q}$ can be figured out by
diagonalizing the tri-diagonal matrices $\left( \mathbf{A-B}\right) \left(
\mathbf{A+B}\right) $ and $\left( \mathbf{A+B}\right) \left( \mathbf{A-B}%
\right) $, which gives rise to $\epsilon_{q}=\sqrt{1+\lambda^{2}+2\lambda%
\cos q}$, i.e., the Eq.~(\ref{HTFdisp}). In Fig.~\ref{Fig_SM2}(a), we
show $\epsilon_{q}$ for various values of $\lambda$. Clearly, $\epsilon_{q}$
is gapless at $\lambda= 1$, and gapped otherwise. Accordingly, the
eigenvectors are given by
\begin{equation}
\begin{aligned} \mathbf{\phi }_{q}&=N_{q}\left( \sin Nq,\sin \left(
N-1\right) q,\ldots ,\sin q\right) ^{T}\text{,}\\ \mathbf{\psi
}_{q}&=N_{q}\left( \sin q,\sin 2q,\ldots ,\sin Nq\right) ^{T}\text{,}
\end{aligned}  \tag{A8}
\end{equation}
where $N_{q}=\left( \sum_{n}\sin nq\right) ^{-1/2}$ is the normalization
factor. The allowed real $q$ are the roots of the equation
\begin{equation}
\frac{\sin\left( N+1\right) q}{\sin Nq}=-\frac{1}{\lambda},  \tag{A9}
\label{TEQ}
\end{equation}
within $q \in[0,\pi]$. In Fig.~\ref{Fig_SM2}(b), we visualize the solutions
to the Eq.~(\ref{TEQ}). The blue solid lines depict the left-hand side
(L.H.S.) of Eq.~\eqref{TEQ}, while the red lines represent its right-hand side (R.H.S.). The
intersections between the two lines correspond to the roots of the equation.
Particularly, for $\lambda\geq1$ [Fig.~\ref{Fig_SM2}(b2) and (b3)], there
are $N$ real roots, corresponding to the bulk modes.

\begin{figure}[t]
\includegraphics[width=0.48\textwidth]{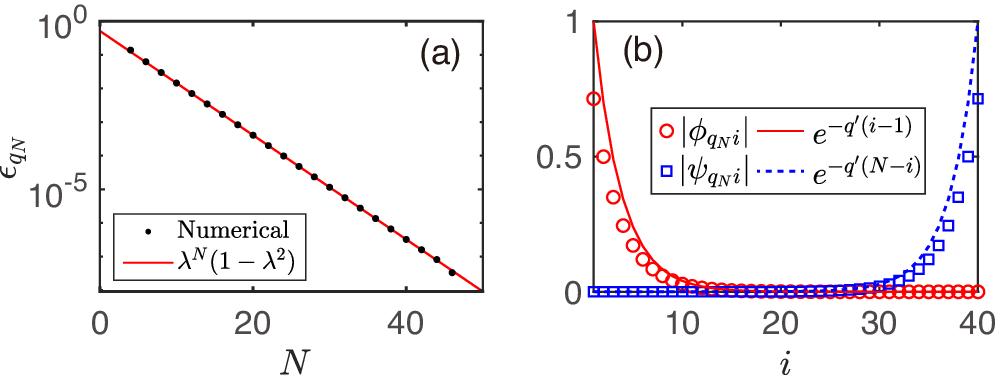}
\caption{(a) Scaling analysis of $\protect\epsilon_{q_{N}}$ on $N$.
Numerical results are represented by dots, while the solid line corresponds
to the asymptotic function $\protect\lambda^{N} (1 - \protect\lambda^{2})$.
(b) $|\protect\phi_{q_{N}i}|$ and $|\protect\psi_{q_{N}i}|$ for the complex-$%
\protect\pi$ mode, with fixed $\protect\lambda=0.7$ and $N=40$. The markers
are results calculated from Eq.~(\protect\ref{Vectorqprime}); solid and
dashed lines are the asymptotic functions $e^{-q^{\prime}(i-1)}$ and $%
e^{-q^{\prime}(N-i)}$, respectively.}
\label{Fig_SM3}
\end{figure}

For $\lambda$ $<1$ [Fig.~\ref{Fig_SM2}(b1)], there exist $N-1$ real roots
and a single imaginary root $q_{N}=\pi+iq^{\prime}$, where $q^{\prime}$ is
determined by
\begin{equation}
\frac{\sinh\left( N+1\right) q^{\prime}}{\sinh Nq^{\prime}}=\frac{1}{\lambda}%
.  \tag{A10}  \label{SM_EQ1}
\end{equation}
Although $q_{N}$ is complex, its corresponding energy $\epsilon_{q_{N}} =
\sqrt{1+\lambda^{2} - 2\lambda\cosh q^{\prime}}$ is real, denoted by an
isolated circle in Fig.~\ref{Fig_SM2}(a). In the large $N$ limit, the
leading term of Eq.~\eqref{SM_EQ1} simplifies to
\begin{equation}
e^{q^{\prime}}=\frac{1}{\lambda},  \tag{A11}
\end{equation}
whose solution is $q^{\prime}=-\log\lambda$. The next-leading approximation
gives $q^{\prime}=\log\left[ {1}/{\lambda}-\left( 1-\lambda^{2}\right)
\lambda^{2N-1}\right] $. Hence, the asymptotic behavior is $%
\epsilon_{q_{N}}\approx\lambda^{N} (1-\lambda^{2})$. In Fig.~\ref{Fig_SM3}(a), we perform a
scaling analysis of $\epsilon_{q_{N}}$ with respect to $N$, where the
numerical results (dots) fit well with the scaling curve (the solid line).

The complex-$\pi$ mode $\eta_{q_{N}}$ is connected to Majorana edge modes in
the following way \cite{K. Chhajed2021,N. Seiberg2023}. The eigenvectors of
the $q_{N}$ mode are expressed as
\begin{equation}
\begin{aligned} \mathbf{\phi }_{q_{N}}&=N_{q^{\prime}}\left( \sinh
Nq^{\prime},-\sinh \left( N-1\right) q^{\prime},\ldots ,- \sinh
q^{\prime}\right) ^{T}\text{,} \\ \mathbf{\psi
}_{q_{N}}&=N_{q^{\prime}}\left( -\sinh q,\sinh 2q^{\prime},\ldots ,\sinh
Nq^{\prime}\right) ^{T}. \end{aligned}  \tag{A12}  \label{Vectorqprime}
\end{equation}
For large $N$, $\sinh(Nq^{\prime}) \approx{e^{Nq^{\prime}}}/{2}$, signifying
that the amplitude of $\phi_{q_{N}}$ decays exponentially from the edge to
the bulk, indicative of edge states. In Fig.~\ref{Fig_SM3}(b), the discrete
markers represent $|\phi_{q_{N}i}|$ and $|\psi_{q_{N}i}|$, clearly
showcasing edge states that align well with the exponential scaling depicted
by the solid lines. In the thermodynamic limit $N \to\infty$, the
eigenvectors tend to be $\mathbf{\phi}_{q_{N}}=\left( 1,0,\ldots,0\right)
^{T}$ and $\mathbf{\psi }_{q_{N}}=\left( 0,0,\ldots,1\right) ^{T}$, leading
to $g_{q_{N}1} = g_{q_{N}N} = \tilde{g}_{q_{N}1} = -\tilde{g}_{q_{N}N} = {1}/%
{2}$. The quasi-particle operators then become
\begin{equation}
\begin{aligned} {\eta}_{q_{N}} & =\frac{1}{2}\left( {c}_{1}+
{c}_{1}^{\dagger}- {c}_{N}^{\dagger}+ {c}_{N}\right) =
\frac{1}{2}(a_{1}-ib_{N}), \\ {\eta}_{q_{N}}^{\dagger} & =\frac{1}{2}\left(
{c}_{1}+ {c}_{1}^{\dagger}- {c}_{N}+ {c}_{N}^{\dagger}\right) =
\frac{1}{2}(a_{1}+ib_{N}), \end{aligned}  \tag{A13}
\end{equation}
where $a_{i}$ and $b_{i}$ are Majorana operators defined as
\begin{equation}
\begin{aligned} a_{i}= {c}_{i}+ {c}_{i}^{\dagger}, \ \ \ b_{i}=i\left(
{c}_{i}- {c} _{i}^{\dagger}\right). \end{aligned}  \tag{A14}
\end{equation}
Hence, the complex mode $\eta_{q_{N}}$ is a linear combination of the
Majorana edge modes $a_{1}$ and $b_{N}$.

\end{document}